# Continuous Transition in Outsourcing: A Case Study

Bilal Raza, Tony Clear and Stephen G. MacDonell
*School of Engineering, Computer and Mathematical Sciences*
*Auckland University of Technology*
*Auckland, New Zealand*
braza@aut.ac.nz, tclear@aut.ac.nz, smacdone@aut.ac.nz

**Abstract**

*Outsourcing is typically considered to occur in three phases: decision, transition and operation. As outsourcing is now well established the switching of vendors and transitioning from one system to another is common. However, most of the research to date on outsourcing has focused on the decision and operation phases, leaving a gap between theory and practice concerning the transition phase. Transition in outsourcing entails the changing of systems, business processes and/or vendors. If a suitable transition approach is not applied pressures for another transition can immediately build. This paper presents results from a case study carried out on the 'Novopay Project' in which the Ministry of Education in New Zealand changed their vendor from an onshore to a near-shore provider. This project resulted in a sequence of three transitions, with each following a different approach as a direct result of the experiences encountered in the previous transition. In this research we made use of the rich 'data dump' of evidence provided by the Ministry of Education (MoE). Our analysis describes how a client organization can become trapped in a continuous transition cycle if a suitable approach is not applied. Transition1 involved the client – MoE – moving from complete outsourcing to selective insourcing. After realizing that they did not have the capabilities to manage insourcing, Transition2 was initiated. In Transition2 the sourcing approach reverted back to complete outsourcing. When it was realized that the new vendor in Transition2 could not in fact deliver a new service model or support end-users in following new business processes, Transition3 was initiated. In Transition3, the client established an internal company to insource service operations to support end-users. Transition can be a sound business strategy initiated for a range of reasons. However, if a flawed sourcing approach is chosen it can result in 'continuous transition'.*

## 1. INTRODUCTION

With the increased use of outsourcing in recent decades there has also been a higher incidence of failures, delays, relationship breakdowns and unsatisfactory performance [1]. The bulk of the outsourcing literature considers it to be a dyadic relationship between client and vendor [2][3], instead of taking into consideration that organizations are now moving towards more selective approaches of combining best-of-breed IT services from multiple vendors [4] [5]. Increasingly, client companies are thus declining to renew or even continue their outsourcing contracts, and therefore switching of vendors (also known as 'transition') is becoming more common [6]. Butler et al. [7] divided the sourcing process into three main phases: *Decision*, *Transition* and *Operation*. Transition follows after contract negotiation and signing and precedes service delivery and operation [1]. In conducting their analysis Butler et al. [7] categorized 116 articles based upon the focus of attention of GSE projects and found that just 2 of those articles were related to the transition phase. This finding coincides with the results of a systematic snapshot mapping study which categorized 301 articles across various dimensions and found that only 19 were related to transition [8]. This suggests that researchers have focused more on establishing the drivers behind outsourcing while less attention has been given to issues relating to the switching of vendors and/or transition of software systems.

Transition is thus a comparatively emergent and under-researched phenomenon of outsourcing, where clients replace their incumbent vendor [9]. It can be assumed that, when clients begin outsourcing for the first time, client-specific knowledge and human resources are transferred from the client organization to the vendor organizations [10]. However, during transition an outgoing vendor has limited interest in supporting an incoming vendor [10]. So what happens when an outsourcing contract expires or when a client organization decides to terminate their current contract with a vendor? [7] The three main options at this stage are either to renegotiate the contract and continue with the same vendor, switch the vendor and continue with outsourcing, or backsource the previously outsourced activities and insource them to an internal entity [11]. In this paper we report the findings of a Novopay case study in which the client became trapped in a cycle of 'continuous transition' wherein they were frequently moving from one sourcing strategy to another. In particular we identify and discuss the main reasons for this continuous transition so as to create better awareness and understanding of how such a situation arises and to possibly help others avoid such circumstances.

The data used in this research was made available under New Zealand's freedom of information laws, chronicled the history and evolution of the Novopay project over several years from its initiation to conclusion. As elaborated in section IV below, the data ranged from cabinet minutes, through project plans, steering committee minutes, correspondence and independent review reports. Although this type of data is not often used in SE related case studies, there are instances in which researchers have employed such data and provided effective results. Verner & Abdullah [12] suggests that secondary data can be valuable for carrying out case studies if there is enough information in the data to answer research questions. However, there is no control by the researcher over the collection of data, and a researcher is constrained by the nature of available data [13]. It can also take a long time to become familiar with such data. To carry out this research, a set of summary Narratives were derived through a key point extraction and coding process, complemented by Antecedent-Consequence diagrams. Narrative analysis provides thick and rich descriptions about the phenomenon, which is considered within a sequence of actions that occurred over time.

Through this set of narratives we illustrate the difficulties encountered, showing how a client may become trapped in a vicious cycle of outsourcing and insourcing. Analysis of rich and complex sequences of events emerging over time in such large and complex undertakings – in this case over a 10-year period – warrant a narrative approach, given its capacity to adequately synthesize and capture the unfolding 'saga' of a process. The remainder of this paper is organized as follows: Section 2 describes the Novopay case, whereas Section 3 and Section 4 describe the method, the type of data used in this research and the approach to data analysis and, respectively. Results are presented in Section 5, with subsequent discussion in Section 6 and limitations in section 7. Section 8 concludes the paper.

## 2. CASE DESCRIPTION

Novopay is the current payroll system of the Ministry of Education, New Zealand. As such it is responsible for paying all teaching and non-teaching staff of the country's public sector schools. It serves around 2500 schools with 120,000 employees, to whom NZ$4.2 billion are disbursed annually. Not only is the payroll system large and complex, it also has some unique sector-specific aspects (e.g., staff may be employed at more than one school at the same time but receive a single payslip; many staff are rehired anew each year). The initial business case for Novopay was approved in 2005. Prior to that the Ministry's payroll system was managed, owned and operated by a New Zealand based vendor, *Datacom* (the outgoing vendor). After the approval of the new business case a full scale tender process was initiated after which *Talent2* (the incoming vendor), an Australian based supplier, was chosen to implement the new payroll system.

During the transition, Novopay went through several variations and delays. An initial setback occurred when the implementation approach was changed after 2 years, which necessitated a second round of tendering. Implementation delays in 2010 led to concerns about the feasibility of the project and the incoming vendor's ability to deliver. As a consequence of this, the project was re-baselined and strengthened at several levels {Education Report: Update on Schools' Payroll Project, Jan 2012}. A new timetable was thus included in the official *variation 1 agreement*: {Project Initiation Contract Part4 Variation Agreement, Oct 2010} with the go-live date set for June 2011. However, there were concerns raised by external quality assurance bodies Independent Quality Assurance New Zealand (IQANZ) (www.iqanz.com) and Price Waterhouse Coopers (PWC) about the timely completion of the project.

The above resulted in a second variation of contract and, subsequently, another timetable was established, with a go-live date of July 2012 {Second Variation Agreement, 2011}. The Novopay system eventually went live in August 2012 with known defects and work-arounds. Subsequently, it faced severe technical issues and major challenges in terms of end-users' acceptance, with significant periods of non-payment or erroneous payment of staff, service shortfalls and general confusion, uproar from the education sector and embarrassment for the Ministry, such that a cabinet minister was appointed to oversee the project. After 2 years of unsatisfactory service in a live production setting, and robust negotiations over non-performance with the vendor, the service operations were back sourced in house by the Ministry of Education (client-MoE) {Education Payroll Limited – Statement of Intent 2015}. The scope of a transition can vary depending upon the context of its project. In the Novopay case, the project involved the changing of a payroll system and vendor for the public education sector. The scope of transition in the Novopay project, and so also for this research, is outlined as follows: *Set up a service desk and payroll centres, Implement a payroll system (or systems), Receive a data extract from the existing payroll system; transform, clean and load the data into the new payroll system(s), Define and implement new service support and service delivery processes, including IT systems, Deliver new business processes, and train Ministry staff and schools payroll support staff* {Project Novopay Review Report, Report by Extrinsic, Jan 2010}.

## 3. RESEARCH METHOD AND APPROACH

In general the need for case study research arises out of a desire to understand a complex phenomenon and to focus on a specific instance or 'case' while retaining an holistic and real-world perspective [14]. Case studies have been increasingly adopted in software engineering in general and in Global Software Engineering in particular – prior research [8] has found that the most dominant methods used in GSE-related research are field and case studies and interviews. Choosing a particular research method implies making a trade-off between *level of control* and *degree of realism*. A case study is ideal for conducting research in a real world setting, permitting a high level of realism, but mostly at the expense of control [15]. Based on the particular research approach, case studies can be classified into four different types: descriptive, exploratory, explanatory or evaluatory [16]. For this research, an exploratory and evaluatory type of case study was considered most suited. The aim of this research was not to test any theory or evaluate any tool, but rather to seek insights and new ideas about the phenomenon of *transition* and to evaluate the feasibility of using a rich mine of available data by applying a plurality of paradigms and data analysis methods.

Reliance on a single case study has often been critiqued for its limited generalizability [17]. However, Yin [14] notes its appropriateness according to five rationales, specifically, when the case is: *critical, unusual, common, revelatory, or longitudinal*. Two of these circumstances arise in this research, *revelatory* and *longitudinal*, meaning that the use of a single case is acceptable. Although switching/transitioning of vendors or software systems is not uncommon in private and public sector organizations, this phenomenon is usually not accessible to researchers. This defines the case under consideration as *revelatory*. The case under consideration, being based on both an initial and an evolving set of data and events, also raised the opportunity to analytically segment and analyze the underlying phenomenon at multiple points in time, thus making it effectively *longitudinal*.

## 4. DATA ANALYSIS USING NARRATIVES

A narrative approach to analysis challenges the commonly prevalent notion that success and failure in systems development are brought about by simple causation [18]. In their simple form narratives provide a way to create shareable understanding of a socio-technical phenomenon [19] in which events and consequences are joined [18]. Davidson [20] demonstrates that narratives can be used to provide deep insights, and Fincham [18] suggests that application of such an approach can be seen as thematic interpretations placed upon a train of events. In other words, narratives present an interpretation of events which constitute core organizational knowledge [21], and these events are made understandable by integrating them within a sequence [18]. Narratives were complemented in this research by antecedent and consequence diagrams as these can help to describe and explain events connected in summary form [22]. They should not be seen as an attempt to create cause-effect type relationships, however. Antecedents are not claimed to be the necessary or sufficient precursors to produce certain consequences.

The Novopay data used for this research was (and is) available in the public domain. This rich mine of data was published by the Ministry of Education, New Zealand on their website, to satisfy official information act stipulations, as well as to head off multiple piecemeal enquiries by the press over the project's many issues. It can be accessed through this web link: *http://www.education.govt.nz/ministry-of-education/information-releases/novopay-information-release*. These types of public 'data dumps' provide an untapped avenue for researchers to consider. Given the potential importance of these data dumps, and noting that there had been limited research focus on the transition phase of outsourcing, when data associated with the Novopay project were released to the public in 2013 we took the opportunity to do a pilot study [23]. After confirming the relevance of the Novopay data from that pilot study, noting the spectacular failure of this transition project, a full-scale data analysis was initiated. Data dumps like this provide an alternate source of data for research. Some researchers even prefer this type of data over surveys or interviews [13]. This data comprised periodic reports from external consulting companies PwC, Deloitte, Equinox, Maven, IQANZ and Extrinsic covering various time periods. These data files are classified in Table I [24]. These covered different phases of the Novopay project from the inception of idea to justification of the project. These included the decision phase, an interim period when transition was going on and the initial operation period after the new vendor took over. Most of these documents were dated between 2008 and 2012-13, but the data corpus also included reports from 1996, 2004-05, 2007 and 2014-15. This provided an opportunity to chronologically arrange data and analyze it longitudinally.

Table I. DATA SOURCES [24]

| Internal Reports and Meeting Minutes | Communication |
|---|---|
| Memos of Ministry of Education (MoE) | Correspondence between representatives of MoE and Exec. of the main vendor |
| Cabinet Meeting Minutes | Emails of end-users |
| Status Reports | **Project Related Documents** |
| Steering Committee Meeting Minutes | Risk Registers |
| Payroll Reference Group Meeting Minutes | Project Initiation Doc. |
| Novopay Board Meeting Minutes | Request For Proposal Doc.- including revised versions |
| Quarterly Meeting Reports for High Risk Projects | Business Case Doc. |
| **Periodic Reports of External Companies** | Fallback plans and proposals |
| | Test Plans and strategies |
| PwC | |
| Deloitte | Communication plans |
| Equinox | Reports about variations in the agreement |
| Maven | Progress review reports |
| IQANZ | End-users Surveys |
| Extrinsic | Remedial plans and programs |

Data analysis primarily followed thematic synthesis steps mentioned in [25]. Thematic synthesis draws on the principles of thematic analysis to infer recurring themes or issues from multiple primary studies. Thematic synthesis steps helped in identifying recurrent themes and build concepts and higher order themes over them. As the Novopay data files had varying levels of data abstraction, the thematic synthesis steps, although time intensive, helped in the movement of understanding constantly from the whole list of recurring themes to its new constituent parts and themes. As the data files were analyzed one by one, understanding of the transition process, both as a whole and its constituent parts, improved. The Novopay data dump comprised 375 PDF files. The data analysis process began with the extraction of key-points from these data-files [24]. The criteria were based on first analyzing and then extracting 'sections of text' if they addressed one or more of the following: stakeholder actions and/or their consequences, stakeholder decisions, risks, issues and concerns. This analysis was conducted by the first author, periodically reviewing and cross checking sets of findings with the other authors.

As most of the data files were directly scanned from physical copies, therefore, the nature of these files were 'image- based'. The downside of working with image-based files is its inability to directly 'select' and 'code' data in NVivo. An initial challenge was to 'code' data in image-based data files. Multiple OCR tools were used to extract text from image based files. These tools had limitations on the number of pages and to overcome this issue, original PDF data files were split using 'Split-PDF' tool and then OCR tools were used on each sub file to extract relevant text. A key-point is a snippet of relevant text, its composition varied from a few lines to paragraphs. It included relevant information either about stakeholders, transition scope or stakeholder actions, which were then used to carry out further analysis. The key-points thus extracted from the data files comprised 724 A4 pages of text, and were afterwards stored in NVivo10 for coding. Coding then resulted in 245 parent and child nodes in NVivo. This process initially followed an inductive reasoning approach, however, after creating a list of codes, it was supplemented by a deductive reasoning process.

The approach taken in this phase was mostly exhaustive rather than selective. This can be indicated from the total 86 parent Nodes of NVivo in which only 12 had one 'reference'. Whereas, the top ten number of references were 95, 77, 65, 62, 55, 48, 48, 46, 43, 42 as indicated in Table II. It indicates the level of saturation achieved in data analysis. After the key-points extraction step, a second level of analysis – step 2 was carried out in which all the key-points were coded in NVivo. In this research, the term 'Coding' is used for this level of analysis. During the coding step, selective and relevant sections of the key-points were identified and labeled across the entire extracted text in key-points.

This Coding process was followed by step 3, 'establishing and categorizing themes'. In this step all the codes were categorized into 'themes', different codes and associated key- points were merged and combined together. Categorized themes were then used in step 4 to explore relationships and connections between themes. During the process of data analysis care was taken to ensure that interpretations were linked to their source by creating a clear chain of evidence and maintaining a link between them. In order to establish trustworthiness and reliability of synthesis in step 5, a researcher lens and external people lens were applied, with the first author as the researcher and the subsequent authors providing the external perspective. These approaches were established and carried out in parallel with all stages of data analysis and synthesis. Afterwards, higher order themes were established. All of the codes and their associated key-points were arranged in a spread-sheet. They were reviewed, re-ordered and combined, resulting in 74 higher order themes (and this paper presents results from a subset of these themes relating particularly to transition). These themes were arranged chronologically in a spread-sheet to support synthesis and the establishment of suitable sequences of themes to enable the development of narratives. These narratives, capturing the initiation and approach for each transition described in Section V, are complemented by graphical representations that also describe the sequence of events.

TABLE II. SAMPLE OF TOP TEN CODES

| Parent Codes | Frequency | Parent Codes | Frequency |
|---|---|---|---|
| Service | 95 | Other risks and issues | 48 |
| Training | 77 | Material Breach | 48 |
| Data Issues | 65 | Resource Constraints | 46 |
| Enduser Issues | 62 | Introduction Novopay | 43 |
| Defects Status | 55 | Transition2 Approach | 42 |

## 5. RESULTS

### A. Transition1:Initiation

This narrative describes the justifications and reasons behind the initiation of Transition1 in the Novopay project. These items are classified into three main categories: client related, vendor related and system related reasons as highlighted in the antecedent-consequence diagram, Fig.1. The argument is supported by the selective use of representative quotes.

The *client-MoE* had been working with an incumbent onshore vendor through a BPO contract for over a decade. The incumbent system was owned, operated, and maintained by this incumbent *Vendor-Datacom*, therefore, the *client-MoE* had limited leverage to negotiate favorable terms and conditions. This heavy reliance contributed towards a 'vendor lock in' situation, as the *client-MoE* had lost expertise to control payroll business processes for the education sector. *"The Ministry's management of the outsourced payroll service provider led to substantial loss of Ministry visibility and knowledge of payroll processes, systems, and performance"* {Novopay Tender Documents, Revised Stage2 Business Case, Nov 2007}. *"Lack of Ministry knowledge placed the incumbent vendor in a monopolist role, with the Ministry uncertain about whether it was getting value for money and whether it had an accurate view of service performance and risk"* {Revised: Stage2 Business Case Payroll Strategy for Schools, Nov 2007}. In addition, over the course of time, changes in policies had also identified the need for end-users to be given direct access to their HR information. It was apparent, however, that the incumbent system could not support this new functionality. *"The lack of access for schools to the payroll system is hindering their ability to effectively manage their employees. Direct access to information would also lower the transaction cost involved with these inquiries, both for the central payroll operation and the schools making the inquiries"* {Stage 2 Business Case, May 2005}.

More generally, the incumbent system was ageing and so only limited features were available in comparison to those offered by modern payroll systems. For example, on-line access; leave management; and linking of salaries to pay grades were highly desired features not available in the incumbent system. It was also anticipated that where new legislation sanctioned changes to processes the incumbent system may not be able to sustain those changes. It would then have to rely on manual and error-prone workarounds. *"Extending the time required to run the existing systems raises the risk of some kind of public failure. This may be through a failure of the payroll itself so that significant number of staff are paid incorrectly or an inability to implement government policy (e.g. tax changes, superannuation)"* {Internal Memo Recommendations for Next Steps, May 2007}. The above contributors urged the client-MoE to move towards Transition1. As legacy systems age their support tends to become more cumbersome. Therefore, vendors develop new versions and transfer their clients. This provides an easier approach for vendors to maintain their systems and also to develop avenues for revenue generation. Any client choosing to continue with the old version may have to bear its full maintenance costs. *"Extending the time of operation on the existing TM4/DATAPAY [Incumbent System] system means that the ministry will be the sole user of TM4/DATAPAY [Incumbent System] for an extended period of time. Datacom is currently migrating its clients to JETPAY [new system] and expects to have completed this before June 2011. The ministry will therefore bear all costs and risks of a functionally and technically obsolete system"* {Internal Memo: Recommendations for Next Steps, May 2007}. *"The projected 'status quo' has also made provision for the increased share of the DATAPAY application overhead to be borne by the Ministry as Datacom's other payroll service customers are migrated off the application leading up to 2011/2012"* {Revised Stage2 Business Case, Nov 2007}. In the Novopay case, the client-MoE could not migrate to a new version because of probity requirements. In the public sector, clients are required to follow an open tender process to choose a new system.

In certain cases, third party vendors cut off support for underlying technologies making it infeasible for the *main-vendor* to keep the incumbent system operational. In Novopay, there was uncertainty about server's operating system and database and whether the third party *vendors* (Microsoft and Oracle) would provide assistance in case of issues. *"The two major sustainability issues identified in this report [Technical review carried out by a third party] are the unsupported versions of the TeacherManager [Incumbent system] Server operating system and database currently used. The Microsoft Windows NT 3.51 operating system used has been unsupported by Microsoft since September 2002, and version 7.2.x of the Oracle Database has been unsupported since October 2000"* {Datacom Payroll Application Sustainability Review, Carried out by Equinox, Aug 2004}. These scenarios encouraged the vendor to push for Transition1.

Automation of business rules helps in consistent implementation of policies, by conceding limited chances for manual interpretations. In Novopay, the incumbent system had limited in-built automated rules which resulted in inconsistent application of policies thus contributing towards Transition1. *"[E]employment agreements are being interpreted manually in different pay centers, resulting in increased risk of business rule inconsistencies and incorrect application of entitlements"* {Cabinet Meeting Minutes, CABMin_05_20_2, June 2005}. Figure 1 below encapsulates these drivers for Transition 1 in an antecedent-consequence diagram.

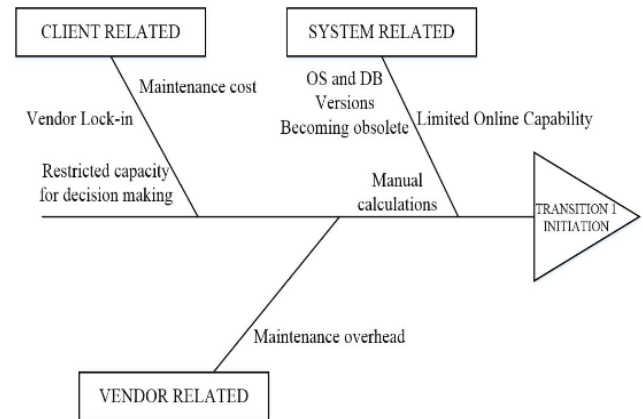

Figure 1. Antecedents which impacted transition initiation

TABLE III. TRANSITION 1 APPROACH

| Selective Insourcing Approach in Transition 1 |
|---|
| The new system will be licensed by the client-MoE |
| The client-MoE will purchase the hardware to operate the system |
| The new system and hardware shall be supported and hosted by third parties |
| Outgoing-Vendor Datacom shall continue to provide operational support services through the new system |
| The client-MoE will increase the number of internal resources to manage the operations effectively |

*B. Transition1:Approach*

Realizing the need for transition had established that continuing the current arrangements carried insurmountable risks. Therefore, it was decided that a new COTS based system be acquired and, afterward, the *client-MoE* integrate its components internally. Hence, a selective insourcing approach was chosen for Transition1. *"This would have involved the ministry purchasing a new payroll system and operating as a 'systems integrator' for all the various components of the solution"* {Cabinet Meeting Minutes, CABMin_08_29_2, July 2008}. *"The ministry was to manage the hosting of the infrastructure, with the pay-clerk services continuing to be provided by the existing supplier until these services could be re-tendered"* {Cabinet Meeting Minutes, CABMin_07_30_3B, Aug 2007}. Subsequently, a request for information (RFI) and request for proposal (RFP) were issued. Some of the key elements of this approach, taken from {Cabinet Meeting Minutes, CABMin_07_30_3B, Aug 2007}, are listed in Table III.

*Selective insourcing* required the *client*–MoE to build internal capabilities. In order to carry this out a special internal unit was set up to help deliver this project. *"The enhancement of the business processes includes the establishment of a Central Advisory Unit (CAU) within the ministry. This would establish a payroll operations and management capability enabling the ministry to manage the complex integration required"* {Revised Stage2 Business Case, Nov 2007}. Using this approach of selective insourcing the two main issues prevalent in the previous service model – loss of Intellectual Property and technical/functional obsolescence – were addressed. *"This approach would address the two key issues of loss of ministry IP and the growing technical and functional* *obsolescence of the DATAPAY/TM4 system"* {Revised Stage2 Business Case, Nov 2007}.

### C. Transition 2: Initiation

The approach taken in Transition1 was subsequently changed during implementation. A revised approach Transition2 was thus considered, only after two years. In this narrative, a brief description is provided about the circumstances which led to this change. One of the central premises for initiating Transition1 was the technical unsustainability of the incumbent system, however, later reviews invalidated this premise. PWC reviewed the earlier business case and concluded that the central premise of 'urgent change' was no longer valid. *"[W]e conclude that the assumptions, upon which the Business Case is based, are no longer valid. Specifically, we consider that the conclusion in the Business Case, that the option of upgrading the existing system presents unacceptably high technology risks, is no longer correct"* {Revised Stage2 Business Case, Nov 2007}.

This advancement overruled *urgent replacement* of the incumbent system and provided an opportunity for the *client-MoE* to reconsider its Transition1 approach. During this time, technology upgrades and new developments, in a combination of platform upgrades and extension of support by the third party IT vendors, extended the potential use of the existing incumbent system. *"One of the key premises for urgent systems change was therefore no longer considered valid"* {Cabinet Meeting Minutes CABMin_07_30_3B, Aug 2007}. In short, the life span for the incumbent system was extended.

Another contributing factor towards changing from the Transition1 approach was the acknowledgement by client-MoE that their capabilities lay in creating policies rather than managing operations of a large scale IT system. *"The operation of systems and processes required by a large payroll is not the core expertise of the ministry"* {Revised Stage2 Business Case, Nov 2007}. Therefore, the client-MoE subsequently reassessed the Transition1 approach and advised the Cabinet that a Business Process Outsourcing (BPO) approach would potentially be more cost-effective and manageable at a lower risk. *"As a result, in Aug 2007 Cabinet rescinded the May 2005 decision for the Ministry to own and operate the schools' payroll system and agreed that the ministry could start the acquisition of a BPO services vendor"* {Novopay Tender Documents – Collated}. Ironically, the Novopay project, which began on a pretext that outsourcing reduces a client's control of their core competencies, was to be outsourced again in Transition2.

### D. Transition 2: Approach

Due to probity requirements of the New Zealand public sector, Transition2 required a complete re-tender of the project i.e., returning to the market and choosing a suitable vendor. One of the concerns about re-tendering was the negative impact it may have on the *client-MoE's* reputation. *"A return to the market will negatively impact the Ministry's reputation as it will be perceived that almost three years have been spent with no decision"* {Internal Memo: Recommendations for Next Steps, May 2007}. *"The market is aware that the ministry has not been able to come to decision for two years and so will be potentially sc[k]eptical regarding their own chances of finalising a contract"* {Internal Memo: Recommendations for Next Steps, May 2007}. Re-tendering the same project may discourage known and capable vendors from further bidding. *"Both vendors have invested significant resources over three years to win the Ministry's business with little perceived progress. Requiring them to continue to invest may result in increased proposed costs to cover the costs incurred to win the Ministry's business. There is also a possibility that Talent2 may withdraw from the process"* {Internal Memo: Recommendations for Next Steps, May 2007}.

An alternative to re-tendering can be to award the new BPO contract to a known vendor, in Novopay it could have been the vendor chosen in Transition1. However this presented a reputational risk with legal repercussions. *"This presents significant reputation risk to the ministry: there is a possibility of third party challenge which would delay implementation, negating the elapsed time advantages, and as this option breaches the Cabinet mandatory procurement guidelines, Cabinet would need to approve this option. This is potential reputation risk to Cabinet"* {Internal Memo: Recommendations for Next Steps, May 2007}.

Another alternative is to carry out a closed tender process between the known vendors. In Novopay, these could have been the vendors chosen for Transition1 i.e., Talent2 and the incumbent vendor-Datacom. However, there were significant probity issues and concerns in following a closed tender process. Therefore, this option was not pursued. *"If a closed tender were conducted between Talent2 and Datacom, the ministry would need to demonstrate reasonable evidence that these were the two most appropriate vendors, and that there were no other vendors able to reasonably deliver the same services to the level required under comparable commercial conditions. Because of the size of the potential contract, the elapsed time since the 2004 tender, and the lack of opportunity for other vendors to bid, there is a possibility of third party challenge and potential delay to implementation"* {Internal Memo: Recommendations for Next Steps, May 2007}.

After considering the various alternatives an open tender process was convened for Transition2; however, it was acknowledged that the most appropriate vendors would either be Talent2 or Datacom. *"Either choosing Talent2 or Datacom has the lowest commercial risk. Costs and conditions have been agreed to a large extent and the competition between Datacom and Talent2 has enabled favourable commercial terms to be gained for the ministry"* {Internal Memo: Recommendations for Next Steps, May 2007}. Drivers for retendering the same project are shown in Figure 2.

As expected, in the retendering process the two main vendors, being the incumbent vendor Datacom and the new vendor Talent2 (chosen for Transition1), became direct competitors. *"Both Talent2 and Datacom were fully aware that they were now in competition with each other"* {Novopay Project History, March 2007}. Talent2 had been the favored choice of the Steering Committee Group in the initial assessment, but Datacom strengthened its position to become a genuine contender. *"With the passage of time the relative position of Datacom has been strengthening and it is not clear that Talent2 would continue as the recommended option in the next draft of the Revised Business Case"* {Novopay Project History, March 2008}.

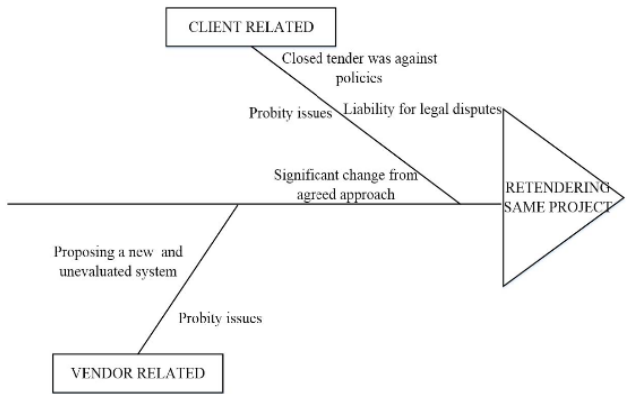

Figure 2. Antecedents which impact on retendering the same project

An unplanned initiation of Transition2 pushed out the project timelines, which, in effect, required the *client-MoE* to extend their Business as Usual (BAU) contract with the incumbent vendor Datacom. In order to negotiate a favorable deal, BAU extensions were worked out before finalizing the vendor for Transition2. As Datacom were bidding for Transition2, it was anticipated, they may extend BAU on conditions favorable to *client-MoE*. However, *"If Datacom is not chosen as the vendor for the new schools' payroll, it will no longer have this incentive. It is therefore important that the contract extension be signed as soon as possible, certainly before the RFP responses from vendors are received and the evaluation starts"* {Revised Stage2 Business Case, Nov 2007}.

Evaluation of the Transition2 proposals found Talent2 fit-for-purpose. Talent2 was therefore selected as the preferred vendor. *"Talent2's proposal had the highest overall weighted evaluation score, could deliver a solution at acceptable risk, and was within the schools' payroll funding envelope"* {Cabinet Meeting Minutes, CABMin_08_29_2, July 2008}. The following Table IV lists the key elements of Transition2.

TABLE IV. TRANSITION 2 APPROACH

| Reverting back to complete outsourcing approach in Transition 2 |
| --- |
| In 2007, Transition1, approach to own and operate the payroll system was cancelled. Acquisition for a BPO vendor was initiated for Transition2 {Novopay Tender Documents – Collated} |
| In Transition2, a single prime vendor will be accountable for a total payroll service {Cabinet Meeting Minutes, CABMin_07_30_3B, Aug 2007} |
| Transition2 will required fewer client resources, still enabling them to retain the expertise required for effective management {Cabinet Meeting Minutes, CABMin_07_30_3B, Aug 2007} |
| Contracts in Transition2 will be carefully constructed to provide strong incentives for the vendor to provide the required quality of services. {Cabinet Meeting Minutes, CABMin_07_30_3B, Aug 2007} |

*E. Transition3: Initiation and Approach*

After the completion of Transition2, the Novopay project went live and the operational phase of outsourcing began. During this time significant issues were faced by end-users. The extent of these issues were such that a cabinet minister was appointed as the 'Minister responsible for Novopay'. It was followed up by a technical review and a Ministerial inquiry [15]. These circumstances compelled the client-MoE to reconsider whether to continue with the same arrangements or carry out another transition, i.e., Transition3. One plausible alternative was to revert back to the old vendor, which might have generated different reactions from various segments of end-users. On the one hand, it could increase the goodwill of end-users struggling with the new system. However it might also have brought disruption for satisfied end-users, as the latter group may not consider another change worthwhile. *"Insufficient change management could result in a rapid loss of any goodwill and potentially significant operational problems"* {Key risk areas, Datacom school's payroll proposal, by Deloitte, April 2013}.

Reverting back to the old vendor would also require challenging and uncertain data cleansing efforts. This was dependent upon coordinated reconciliation activities to resolve any data related issues. *"Reconciliation will be difficult and will take a long time"* {Outlining Proposal for Resuming Payroll Services, by Datacom, March 2013}. *"Some data may be impossible to reconcile"* {Outlining Proposal for Resuming Payroll Services, by Datacom, March 2013}. After the completion of a transition the operations are taken over by a new vendor and the old vendor may not have access to or visibility of the data stored in the new system. At this stage, if a client decides to revert back to the old vendor (i.e., Transition3) then it would require another cycle of transition and support from the new vendor. *"All of these will require support from Talent2. The approach to data conversion will require significant effort from schools"* {Key risk areas, Datacom school's payroll proposal, by Deloitte, April 2013}. *"Inadequate management of data activities could increase project costs and result in significant payroll issues"* {Key risk areas, Datacom school's payroll proposal, by Deloitte, April 2013}.

In the Novopay case, it was concluded that another transition would bring significant disruption to the sector again. *"This would complicate the management and governance demands on the Ministry. The underlying commercial and contractual arrangements would be complex"* {Key risk areas, Datacom school's payroll proposal, by Deloitte, April 2013}. *"In addition, there would be challenges in keeping the Novopay service operating during the transition. It is likely that Novopay would lose key people and there would be continuing contention between Novopay, Datacom, and the Ministry over the relatively small pool of people with deep schools payroll expertise"* {Education Report: Recommendations on Datacom contingency, April 2013}. Thus, it was decided to continue with the new vendor, *"the Ministry is recommending that the current proposed contingency option not be taken up"* {Education Report: Recommendations on Datacom contingency, April 2013}.

However, after approximately two years, when end-users continued to face issues and continued to complain about the level of service, the client-MoE decided to set up an internal crown entity Education Payroll Limited (EPL) to take over operational responsibilities from the vendor Talent2. Thus Transition3 was initiated. Operations related to processing of payroll were then backsourced to EPL.

*"EPL is a Crown company established to provide payroll services to New Zealand's schools. We were incorporated in August 2014 to take over the operation of the schools payroll service from Talent2 in October 2014. Ownership is held equally between two Shareholding Ministers, the Minister Responsible for Novopay and the Minister of Finance. EPL is governed by a Crown appointed Board of Directors"* {Education Payroll Limited, Statement of Intent, 2015}. Setting up an internal entity required another transition to ensure that the new set up has the capabilities to operate and the services are successfully transferred. Table V lists the key elements of Transition3.

The following Figure 3 highlights this phenomenon of volatility in sourcing strategies and continuous transition.

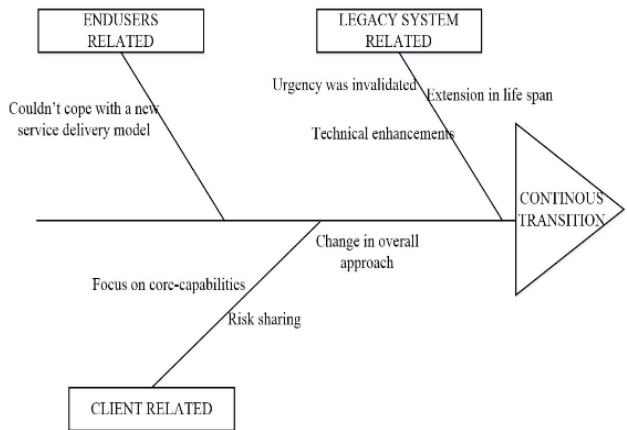

Figure 3. Antecedents which impact on continuing transition

TABLE V. TRANSITION 3 APPROACH

| Moving payroll service to a Government owned company in Transition 3 |
|---|
| Internal company EPL was incorporated in August 2014 for Transition3 to take over operational services from the vendor chosen in Transition2 {Education Payroll Limited – Statement of Intent 2015} |
| Ownership of this internal company was held between two Shareholding Ministers, the Minister Responsible for Novopay and the Minister of Finance. It was governed by a Crown appointed Board of Directors {Education Payroll Limited – Statement of Intent 2015} |
| Transition3 changed the previous approach by shifting from decentralized service delivery model to a centralized service delivery model. |
| EPL was responsible for payroll processing, customer service functions and all operational relationships. {Education Payroll Limited – Statement of Intent 2015} |

## 6. DISCUSSION

Continuous and excessive reliance upon the same vendor in outsourcing may cause loss of knowledge and vendor lock-in. Deventer & Singh [26] mentioned two schemes which vendors can use to effect this: *'hard assets'* and *'soft assets'*. Use of proprietary hardware and software is classified as a hard asset, whereas relationship building to heighten switching costs is classified as a soft asset. The results of this study also support this classification. In the Novopay project, the incumbent vendor developed personal relationships with end-users, so that end-users were excessively reliant upon them to carry out operations. The system was also owned and operated by the vendor. However, a prolonged vendor lock-in situation may compel a client to carry out a transition, due to loss of knowledge and the desire to regain control. The findings of this study suggest that transition can be initiated by a client, vendor or due to limitations of the incumbent system. In certain situations, it becomes inevitable to carry-out a transition. Over a period of time, vendors may start migrating clients to a new version of their applications to manage support more easily. Legacy systems can also become unsupportable due to ICT advancements and changes in underlying technologies. Moreover, in public sector projects, changes in legislation can create pressures for further customizations which can be risky to implement in legacy systems. So, at times, avoiding *transition* could itself be risky and infeasible.

From the results of this study, it was found that there are two main reasons for an incumbent system to become obsolete. Underlying obsolete technologies, e.g., in this case the payroll system was written in an older version of COBOL, finding experts for which could become difficult. Secondly, third party companies may stop support of their older versions of Operating Systems (OS) and Database Management Systems (DBMS) making it difficult for vendors to keep providing support indefinitely. These aspects should be considered and factored in while negotiating contracts. Vendors can develop partnerships with third party vendors to implement simpler tools and processes for shifting customers to their new platforms or versions with ease. If a COTS based system is heavily customized for a specific client, then vendors may find it difficult to shift it to their new version. Therefore, it is imperative to keep track of the customized version to ensure that a smooth process is developed. Results of this study shows that if knowledge about customizations is not properly documented and kept up to date then it may cause over-reliance on specific human resources. Due to this reason an incoming vendor can be excessively reliant upon the outgoing vendor's support for transition. According to Mirani [27], an outsourcing relationship is not fixed, rather it is variable and changes with time. It was elaborated that clients often initiate their outsourcing relationship by using simpler and straight forward tasks. After establishing an initial contract and following the course of time, clients begin to assign complex applications to selected vendors which necessitates establishing a more loose, trust and network based relationship. After following this latter approach, over a period of time, applications become business critical. Clients may then establish a command-based hierarchy to gain more control over their critical operations. The latter necessitates either acquiring some form of formal stake or setting up some form of subsidiary of its own [27]. The results of this study have similarities and contrasts with the Mirani [27] framework as shown in Figure 4. To begin with, the client-MoE already had a business critical application outsourced to an external vendor. In order to regain control from the vendor it was decided that *Transition1* should be carried out, in which the client will insource the management of the implementation (of a new system). Changing the sourcing strategy requires changing the client's roles. Yakhlef & Sié [28] described the process of moving from insourcing to outsourcing as one of producer to purchaser of services and the type of knowledge which a client requires to sustain this change. In contrast, the Novopay *Transition1* followed the process of changing from purchaser to provider of services – outsourcing to insourcing. Insourcing requires reintegration of knowledge, developing new capabilities and competencies [29]. Building such capabilities internally for a public sector agency may not be a straight forward endeavor. It may also not align with their strategic goals.

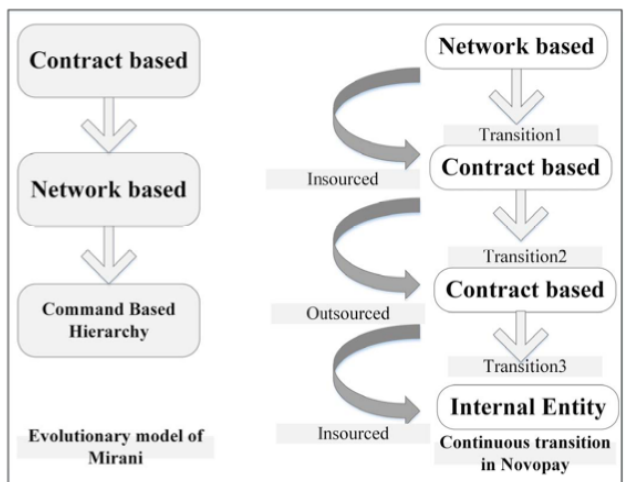

Figure 4. Comparison between Mirani's Evolutionary Model and Continuous Transition in Novopay

In the Queensland Health payroll project, when deciding about the implementation approach, a similar comparison was made between setting up an internal entity (insourcing) and choosing an external vendor (outsourcing). According to Queensland Health Payroll System Commission of Inquiry Report there were contrary views about this strategic choice. Mr. Goddard (Project Management Consultant) and Mr. Uhlmann (Expert in IT Projects Delivery) were of the view that an external vendor may not lower the cost of the overall project but certainly increase the risk of government losing control of its own program. Mr. Uhlmann's statement as recorded in the Queensland Health Payroll System Commission of Inquiry Report is worth noting here: *"You could have brought someone in to ... to bring all the project disciplines into play, get all the right people with the right sort of expertise supporting around the PMO ... get that applied to your current partners and then drive that and hold them accountable ... . If ... the prime contractor's role is ... to replace all of that ... I would not have support (sic) that ... [because] ... it's about who can best accelerate the packages of work ... and ... whoever has got the background knowledge and the skills and expertise on the ground, you want to leverage that ... you would not get rid of that sort of knowledge and background expertise ..."* {Queensland Health Payroll System Commission of Inquiry Report, 2013, pp. 85}. These views of Mr. Goddard and Mr. Uhlmann from the QH payroll project coincide with the actions and justifications for the decision taken in transition 1 of Novopay. It was argued that payroll is business critical and the client had lost knowledge and control of it – but MoE didn't have a choice to create a command-based hierarchy (by taking ownership of the existing supplier) and therefore, opted for an insourcing approach. However, an insourcing approach requires building up internal capabilities. External vendors benefit when services are outsourced to them and they may make diligent efforts to transfer knowledge. In insourcing, vendors may not have the same level of enthusiasm and motivation to help manage the transfer of knowledge back to the client. Lack of preparedness and lack of internal capabilities to manage insourcing may trigger another sourcing cycle - *Transition2* in the case of Novopay. After approximately two years, following Transition1 (an insourcing strategy), the client determined they lacked the capabilities to manage payroll operations internally. It was then decided to re-model the approach and follow a complete business process outsourcing, which would help share risks and reduce cost for overall operations. Thus, *Transition2* was initiated. Outsourcing ostensibly enables client organizations to share risks with vendors. However, the level of service that end users expect may not be provided through a new vendor. Clients are thus required to build the confidence and capacities of end-users and provide support and leverage in getting them settled with a new vendor. External vendors may have limited interest in building up end-user capabilities, if it is not previously arranged. In the Novopay project, end-users faced major issues with service quality after the *Transition2* was completed. The client MoE had to eventually create an internal entity to manage the service delivery and insource the operational activities in *Transition3*. Comparison between Mirani's evolutionary model and the continuous transition phenomenon evident in Novopay is shown in Figure 4. Moving from a *network based* relationship to a *contract based* relationship in Transition1 with known roles may have been viable, but the more ambitious outsourcing in *Transition2* probably needed a *network based* relationship, given the inability to fully contractually specify the services to be delivered. In Transition 3 the MoE established an internal entity which was effectively a *command based hierarchy* in order to regain control.

## 7. LIMITATIONS

The results of this paper are based on a single case study. However, the case study is analysed *longitudinally* (over a 10- year duration) and *revealing* which justify using a single case study. Due to the controversial nature of the transition phenomenon there can be difficulties in accessing relevant data. This makes Novopay a revealing case. Moreover, the data which has been used in this research relates to different time frames which provided an opportunity to segment and analyze data at multiple times making the analysis longitudinal and unique.

Another limitation concerns the nature of the data used in this study. Novopay data was provided by the Ministry of Education, New Zealand, due to the Official Information Act 1982. The data dump was not collected for the purpose of carrying out research and the researchers had limited control over it. However, this limitation increased the objectivity of the data and reduced the researchers' bias in data collection. On the downside, it took more time to establish connections between concepts within the voluminous data. Data dumps like Novopay provide another avenue and possibly untapped data sources for software engineering researchers to explore. As the analysis was conducted as part of a doctoral study, the coding was largely conducted by a single rater, apart from a pilot study conducted in collaboration between the three authors. However findings were regularly cross-checked with the supervisors of the study, for plausibility and comprehensiveness. The use of a clearly defined protocol for the thematic analysis further helped build rigor into the study. The results of this research are more generalizable to a public-private partnership (PPP) context as can be noticed from the similarities with the Queensland Health (QH) Payroll project. However, a fuller analysis of the QH payroll project is still to be carried out. This is being considered as a future development in this research.

## 8. CONCLUSION

This paper presents narratives revolving around continuous transition of the Novopay case study. It involved changing payroll system, service delivery model and refining business processes. It also included switching from an onshore to a near-shore vendor. The Novopay project, unlike originally planned, went through three distinct transitions.

- Transition1 followed an insourcing approach to address *loss of knowledge* and *control* of client-MoE over the payroll operations. However, with more control comes more responsibilities and risks, which the client-MoE was not prepared to manage. Insourcing was also not considered in-line with core competencies of client-MoE. The above paved the way for another transition.

- In Transition2, loss of control and knowledge through external vendor, the premises on which Transition1 was initiated, was overruled for sharing risks with an external vendor. Thus the project was completely outsourced again. However the effect on end-users of vendor nonperformance was not carefully considered.

- The third transition, Transition3 was initiated due to end-users' dissatisfaction about following new business processes and concerns about working with a new vendor. The new business processes did not effectively service end-users. The client then established an internal entity to decentralize operational services to help support end-users in Transition3.

This shows that the process of Transition can carry forward and propagate to another cycle if clients do not choose the right sourcing model or strategy at the relevant point in time. Understanding internal capabilities and acknowledging the varying capabilities of vendors and end-users are imperative in this regard. Clients should develop multivariate strategies to manage incumbent vendors whilst supporting end-users and keeping track of the obsolescence of incumbent-systems to become transition-ready. It must be noted that a network-based type of relationship cannot be established instantaneously. Rather, they are developed progressively. Therefore, public sector clients should be conscious of the relationship between end-users and vendors.